\documentclass[conference]{IEEEtran}
\IEEEoverridecommandlockouts

\usepackage{cite}
\usepackage{amsmath,amssymb,amsthm}
\usepackage{graphicx}
\usepackage{textcomp}
\usepackage{xcolor}
\usepackage{booktabs}
\usepackage{multirow}
\usepackage{subcaption}
\usepackage{hyperref}
\hypersetup{hidelinks}
\usepackage{url}

\begin{document}

\title{Evaluating Container Orchestration for Neuromorphic Workloads in Virtual Edge Environments}

\author{
  \IEEEauthorblockN{Huyen Pham, Bilhanan Silverajan}
}

\maketitle

\begin{abstract}
The growing adoption of edge computing has created an increasing need for workloads capable of operating under strict resource and energy constraints. Neuromorphic computing, and spiking neural networks (SNNs) in particular, offers an energy-efficient alternative to conventional machine learning through event-driven computation. However, how SNN workloads behave when deployed within modern container orchestration frameworks, especially in edge environments, remains largely unexplored. This paper investigates the feasibility of deploying and orchestrating SNN workloads in a virtual edge environment using Kubernetes, focusing on end-to-end latency, throughput, classification accuracy, infrastructure overhead, and runtime behavior under concurrent load.

Experiments were conducted on a single-node K3d cluster running on a Windows 11 host with WSL2 and Docker Desktop. The results show that SNN workloads are highly sensitive to resource availability. Restricting CPU to 0.5 cores increased median latency by 47.6$\times$ and reduced throughput by 49$\times$, while the most constrained configuration failed entirely due to insufficient memory. Classification accuracy remained stable across all working configurations. From an orchestration perspective, K3d successfully deployed and scaled SNN workloads, though its default round-robin routing policy introduced severe tail latency under replica scaling, highlighting a mismatch between stateless load balancing assumptions and long-running inference workloads.

Overall, the study provides a practical baseline for deploying neuromorphic workloads in containerized edge environments and highlights the importance of resource provisioning and orchestration configuration. Future work should explore improved routing strategies, memory optimization, and validation on physical edge hardware.
\end{abstract}


\section{Introduction}

With the rapid expansion of edge computing environments, there exists an increasing need for workloads capable of operating within strict resource constraints, limited power budgets, and variable network connectivity. Traditional machine learning approaches are often computationally intensive and cloud-dependent, making them less suitable for deployment on constrained edge devices~\cite{rossi_neuro-inspired_2025,deng_edge_2025}. Neuromorphic computing offers an alternative architectural solution, drawing inspiration from biological neural systems to enable energy-efficient, event-driven computation through time-dependent spike-based signalling.

Spiking neural networks (SNNs) embody these principles in software, enabling neuro-inspired computation without requiring dedicated hardware. Unlike artificial neural networks (ANNs), SNNs are sparse, asynchronous, and inherently temporal. As SNN-based applications see growing adoption in real-world use cases, the question of how to reliably deploy and manage them across distributed infrastructure becomes increasingly relevant.

Modern orchestration technologies such as Kubernetes are well known for their automation, scalability, and resource management of distributed workloads~\cite{das_container_2025}. The rise in edge computing has introduced lightweight Kubernetes distributions such as K3s to support deployment in resource-constrained environments. K3d was developed as a wrapper for K3s, allowing multi-node clusters to run within Docker containers on a single host. However, these frameworks are typically designed for workloads with stable resource usage and continuous execution patterns, commonly found in conventional cloud-native applications. This raises an important question as to whether current orchestration frameworks are compatible with the irregular, event-driven execution patterns of SNN workloads.

This paper addresses this gap by investigating how simulated SNN workloads behave when deployed in containerized edge environments, focusing on a virtualized edge infrastructure on a single Windows host. Two research questions guide the study:

\begin{itemize}
  \item \textbf{RQ1:} How does inference performance (latency and throughput) of an SNN workload degrade as container resource constraints are tightened to simulate edge conditions?
  \item \textbf{RQ2:} How does K3d perform in terms of deployment complexity and runtime behaviour when managing SNN inference workloads?
\end{itemize}

The full implementation, including source code, Kubernetes manifests, and benchmark scripts, is publicly available to ensure reproducibility~\cite{pham_jennyhphamdistributed_neuro_sim_2026}.

\section{Background and Related Work}

\subsection{Neuromorphic Computing and SNNs}

Neuromorphic computing is a computational paradigm inspired by the structure and behavior of the human brain~\cite{kudithipudi_neuromorphic_2025}. Unlike conventional von Neumann architectures, neuromorphic systems reduces data movement overhead by bringing memory and computation closer together~\cite{shrestha_survey_2022}. 

A key feature of neuromorphic systems is their use of spike-based communication. Instead of relying on continuous numerical values, neurons are activated through discrete spike events. This give rise to benefits such as sparse, asynchronous, and event-driven computation~\cite{rathi_exploring_2023}.

The underlying principle of neuromorphic computing systems is modelled through spiking neural networks (SNNs). Information in SNNs is encoded when a spike occurs and whether it occurs at all \cite{rathi_exploring_2023}. This temporal information allows SNNs to capture patterns in the timing of inputs, supporting more efficient and event-driven computation ~\cite{kim_exploring_2022,rossi_neuro-inspired_2025}. The spiking pattern of each neuron can be described mathematically through the Leaky Integrate-and-Fire (LIF) model~\cite{hunsberger_spiking_2015,rathi_exploring_2023}:
\begin{equation}
\tau_{RC}\frac{dv(t)}{dt} = -v(t) + RI(t),
\label{lif}
\end{equation}
where $\tau_{RC}$ is the membrane time constant, $R$ is the membrane resistance, and $I(t)$ is the input current. A neuron’s membrane potential $v(t)$ accumulate incoming signals $RI(t)$ over time while also gradually discharging. It will fire a spike to other connected neurons once the threshold $V_{th}$ is reached, and then becomes unavailable for a refractory period $\tau _{ref}$. After this period, the neuron resumes its normal functionality as described in Equation \eqref{lif}.

In this paper, spike-timing-dependent plasticity (STDP) is used as the learning mechanism. STDP is an unsupervised learning rule based on Hebbian learning, where synaptic weights between connected neurons are adjusted according to the relative timing of pre- and post-synaptic spikes ~\cite{rathi_exploring_2023,yamazaki_spiking_2022}. This is combined with lateral inhibition, where the activation of a neuron will shortly prevent its neighboring neurons from firing ~\cite{yi_learning_2023,dong_unsupervised_2023}. 

Well-known neuromorphic hardware platforms include IBM TrueNorth, Intel Loihi, and SpiNNaker2~\cite{debole_truenorth_2019,davies_loihi_2018,mayr_spinnaker_2019}. Since such hardware remains limited, this work uses software simulation via BindsNET on conventional CPU~\cite{hazan_bindsnet_2018}.

\subsection{Edge Computing and Container Orchestration}

Edge computing refers to the process by which computations are performed near data sources rather than in centralized cloud platforms ~\cite{shi_edge_2016,cao_overview_2020}.  By keeping computation closer to the source, edge devices achieve faster response times, lower latency, and reduced cloud reliance. However, edge nodes are typically constrained by limited CPU, memory, and energy budgets~\cite{p_comprehensive_2025}. A  virtual edge environments simulates these characteristics using container-based technologies on a single host~\cite{qi_survey_2025}, making it possible to simulate consistent and reproducible experimental conditions without hardware~\cite{symeonides_fogify_2020}.

Kubernetes is commonly used to automate the deployment, scheduling, and scaling of containerized workloads. It works by abstracting containers into pods, which are then executed on cluster nodes based on their CPU and memory requirements~\cite{burns_borg_2016}. Lightweight distributions such as K3s reduces the Kubernetes memory footprint to as little as 512 MB while retaining core orchestration capabilities~\cite{das_container_2025}.  K3d builds on this by running a complete K3s cluster
entirely within Docker containers on a single host. 

However, Kubernetes distributions were designed for workloads with predictable and continuous resource usage. SNN workloads do not fit this profile,
as their computational patterns and resource usage are irregular and bursty~\cite{chen_skydiver_2022}. As a result, standard scheduling, auto-scaling, and health-check mechanisms may misinterpret periods of low spiking activity or fail to account for varying energy bursts.

\subsection{Related Work}

Deng et al.~\cite{deng_edge_2025} provide a comprehensive survey on edge intelligence based
on SNNs and examines current techniques for adapting SNNs to edge constraints.  Their work highlights the potential of SNNs to reduce latency, bandwidth usage, and power consumption on constrained devices, while noting that evaluating SNNs on conventional hardware does not fully reflect their capabilities. Xue et al.~\cite{xue_edgemap_2023} further examined this challenge in the context of deployment, highlighting that deploying SNNs on edge devices remains highly complex due to temporal dynamics, spike-based communication, and limited resources. Rossi et al.~\cite{rossi_neuro-inspired_2025} demonstrate how utilizing SNNs for task offloading decisions in edge-IoT networks can improve latency and energy efficiency. However, in these cases, SNNs are
used to optimize orchestration processes rather than being deployed as workloads.

Regarding orchestration, Cili\v{c} et al.~\cite{cilic_performance_2023} evaluate Kubernetes and its lightweight variants, finding that K3s and KubeEdge introduce overhead affecting deployment and scheduling. Xu et al.~\cite{xu_performance_2017} conclude that containerization introduces negligible overhead for conventional deep learning workloads. More recently, DeBole et al.~\cite{debole_scalable_2025} introduce a system implementing containerized neural inference workloads on NorthPole hardware with high energy efficiency. However, the paper focuses on specialized hardware rather than edge environments, and utilizes conventional networks rather than SNN workloads.

Across these works, the interaction between neuromorphic-inspired workloads and modern container orchestration platforms remains largely unexplored. This paper aims to address that gap, thus serving as a baseline for future work on neuromorphic workload orchestration at the edge.

\section{Methodology}

\subsection{System Overview}

\begin{figure}[h]
  \centering
  \includegraphics[width=\linewidth]{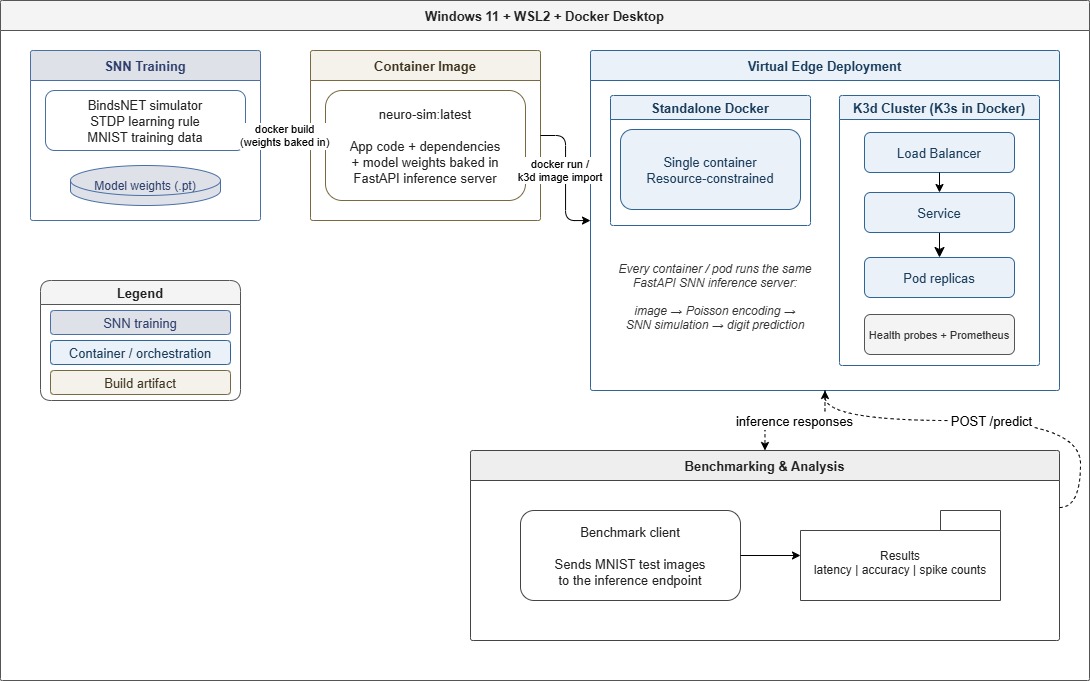}
  \caption{System overview illustrating training, containerization, deployment, and benchmarking phases.}
  \label{fig:system_overview}
\end{figure}

The system runs on a single Windows 11 host using WSL2 and Docker Desktop, and is structured into four phases: model training, containerization, deployment, and benchmarking (Figure~\ref{fig:system_overview}). Initially, an SNN model is trained offline using the full MNIST dataset (60,000 training and 10,000 test samples across ten digit classes~\cite{noauthor_mnist_nodate}), after which all learned weights are fixed. This ensures that subsequent experiments reflect inference behavior exclusively.

The model weights are incorporated into a Docker image together with the application code and a FastAPI inference server. The application is evaluated across two deployment contexts. In the first scenario, the image is run as a standalone Docker container under varying CPU and memory constraints. In the second scenario, the same image is deployed within a K3d cluster to evaluate orchestration behavior,

\subsection{SNN Model}

The BindsNET Python framework~\cite{hazan_bindsnet_2018} was selected for its built-in STDP support and PyTorch backend. Model training was performed with GPU acceleration via CUDA, while inference was ran exclusively on CPU to ensure consistent evaluation and reflect containerized edge constraints.

\begin{figure}[htbp]
  \centering
  \includegraphics[width=0.85\linewidth]{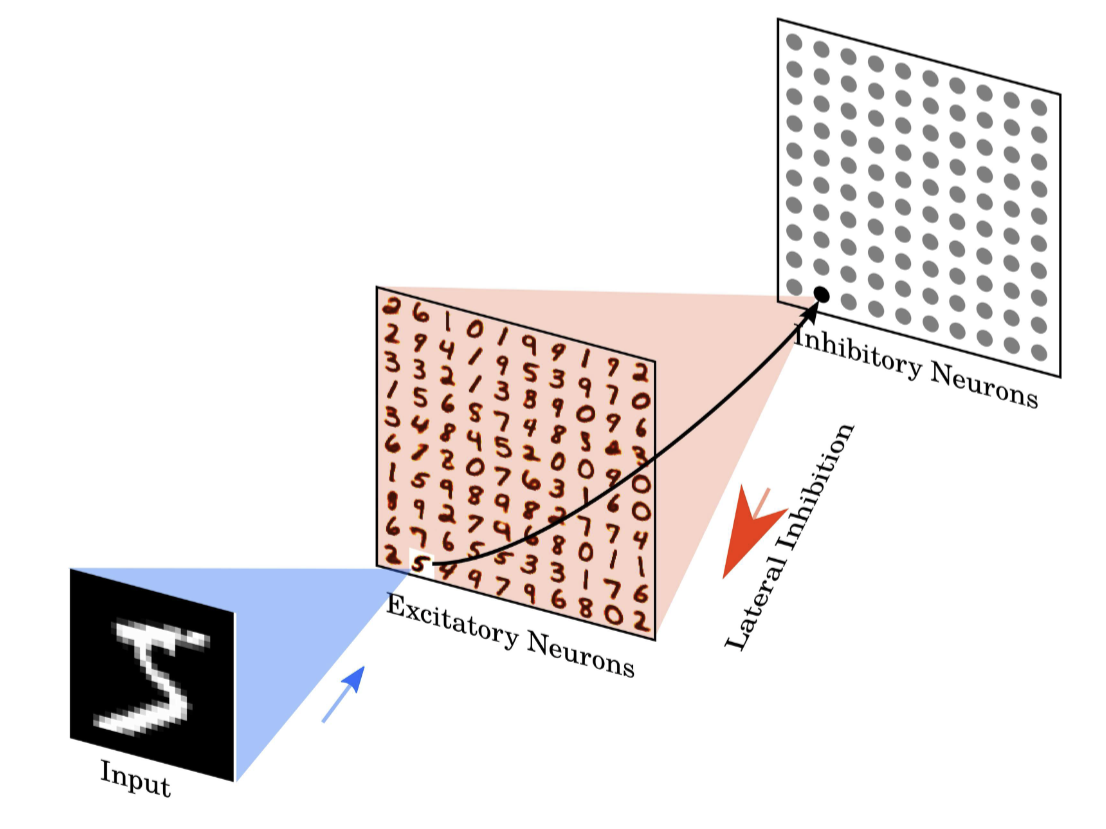}
  \caption{Network architecture of the Diehl and Cook SNN model~\cite{diehl_unsupervised_2015}.}
  \label{fig:diehl_arch}
\end{figure}

Diehl and Cook's unsupervised digit recognition model~\cite{diehl_unsupervised_2015} was used as the network architecture (Figure~\ref{fig:diehl_arch}). It was replicated with near accuracy by BindsNet, with the addition of a Self-Organizing Map (SOM) property for digit clustering. The resulting network consists of an input layer of 784 neurons, an excitatory population of 400 neurons, and a corresponding inhibitory layer that implements lateral inhibition. The key parameters of the model and configurations are
compiled in Table~\ref{tab:model_params}.

\begin{table}[htbp]
\centering
\caption{Key parameters of the SNN model and training configuration}
\label{tab:model_params}
\begin{tabular}{lll}
\toprule
\textbf{Category} & \textbf{Parameter} & \textbf{Value} \\
\midrule
\multirow{3}{*}{Architecture} 
 & Excitatory neurons & 400 \\
 & Input neurons & 784 ($28 \times 28$) \\
 & Simulation timestep & 1.0 ms \\
\midrule
\multirow{3}{*}{Training} 
 & Training samples & 60{,}000 \\
 & Batch size & 1 \\
 & Input intensity & 64.0 \\
\midrule
\multirow{3}{*}{Lateral Inhibition} 
 & Initial inhibition & 10.0 \\
 & Maximum inhibition & $-40.0$ \\
 & Inhibition update interval & every 500 samples \\
\midrule
\multirow{2}{*}{Inference} 
 & Simulation window & 100 ms \\
 & Test samples & 10{,}000 \\
\bottomrule
\end{tabular}
\end{table}

During inference, each input image is converted into spike-based representations using Poisson encoding and passed to the network over a fixed simulation window of 100 ms. Network output is decoded using two methods. The first is an \textit{all-activity} approach, where the predicted class corresponds to the neuron group with the highest total spike activity. The second is a  \textit{proportion-weighting} method, where spike
activity from all neurons is aggregated and grouped by their class labels. The predicted class will be determined based on the relative distribution of this activity.

\subsection{Deployment Configurations}

The inference service is packaged as a Docker image using \texttt{python:3.11-slim}, including the FastAPI application, a CPU-only PyTorch build, BindsNET, and pre-baked model weights. For RQ1, three standalone Docker configurations are evaluated:
\begin{itemize}
  \item \textbf{Baseline}: no CPU or memory limits
  \item \textbf{Constrained 1}: 0.5 CPU cores, 512 MB memory
  \item \textbf{Constrained 2}: 0.25 CPU cores, 256 MB memory
\end{itemize}

The 0.5 CPU/ 512 MB setup was chosen as it meets the minimum memory requirements of lightweight Kubernetes distributions~\cite{das_container_2025}. The 0.25 CPU / 256 MB setup falls deliberately below this threshold to define a failure boundary.  In both constrained environments, memory swap is disabled to ensure that performance reflects strict resource availability and is hence consistent with edge environments

For RQ2, the same image is deployed in a single-node K3d cluster with pod resource requests of 500m CPU and 512Mi memory, and limits of 1000m CPU and 1Gi memory. Three concurrent load scenarios are evaluated: (1) 1 replica, 1 client; (2) 1 replica, 3 clients; and (3) 3 replicas, 3 clients, each sending 50 requests.

\subsection{Evaluation Metrics}

\textbf{End-to-end latency} is the primary performance metric,  measured as the elapsed time in milliseconds from when a request is sent to when the response is received. Results are reported at p50, p75, p95, p99, and maximum. For RQ1, p95 is used as the primary tail metric since latency is relatively stable. For RQ2, p99 is used instead to capture worst-case queuing effects caused by load balancing

\textbf{Infrastructure overhead} is calculated from the difference between the p50 latency and the minimum observed latency. This metric approximates the overhead introduced by the container runtime, networking stack, and API framework in addition to the baseline cost of the SNN simulation.

\textbf{Throughput} is the number of successfully completed requests per second over each benchmark run.

\textbf{Classification accuracy} is evaluated using both decoding methods across all MNIST digit classes, assessing whether resource constraints affect prediction quality.

\textbf{Spike count} records the total number of output-layer spikes generated during the simulation window. It is used as a proxy for workload intensity, where higher spike activity
reflects greater computational effort within the network.

\textbf{Failure conditions} are captured at the scenario level, including container OOM termination and startup timeouts.

\subsection{Experimental Environment}

All experiments were carried out on a single Lenovo Yoga Pro 9 16IAH10 laptop (Intel Core Ultra 9 285H, 64 GB RAM, Windows 11) kept plugged in throughout to avoid performance variability. Software versions are listed in Table~\ref{tab:versions}.

\renewcommand{\arraystretch}{1.3}
\setlength{\tabcolsep}{8pt}

\begin{table}[htbp]
\centering
\caption{Software versions used across application and infrastructure layers}
\label{tab:versions}
\begin{tabular}{|l|l|}
\hline
\textbf{Component} & \textbf{Version} \\
\hline
Python & 3.12.3 \\\hline
PyTorch (CPU-only) & 2.9.1 \\\hline
BindsNET & 0.3.3 \\\hline
FastAPI & 0.135.1 \\\hline
Docker & 28.2.2 \\\hline
K3d & v5.8.3 \\\hline
K3s & v1.31.5-k3s1 \\
\hline
\end{tabular}
\end{table}

\section{Results and Discussion}

\subsection{RQ1: Inference Under Resource Constraints}

\begin{figure}[htbp]
  \centering
  \includegraphics[width=\linewidth]{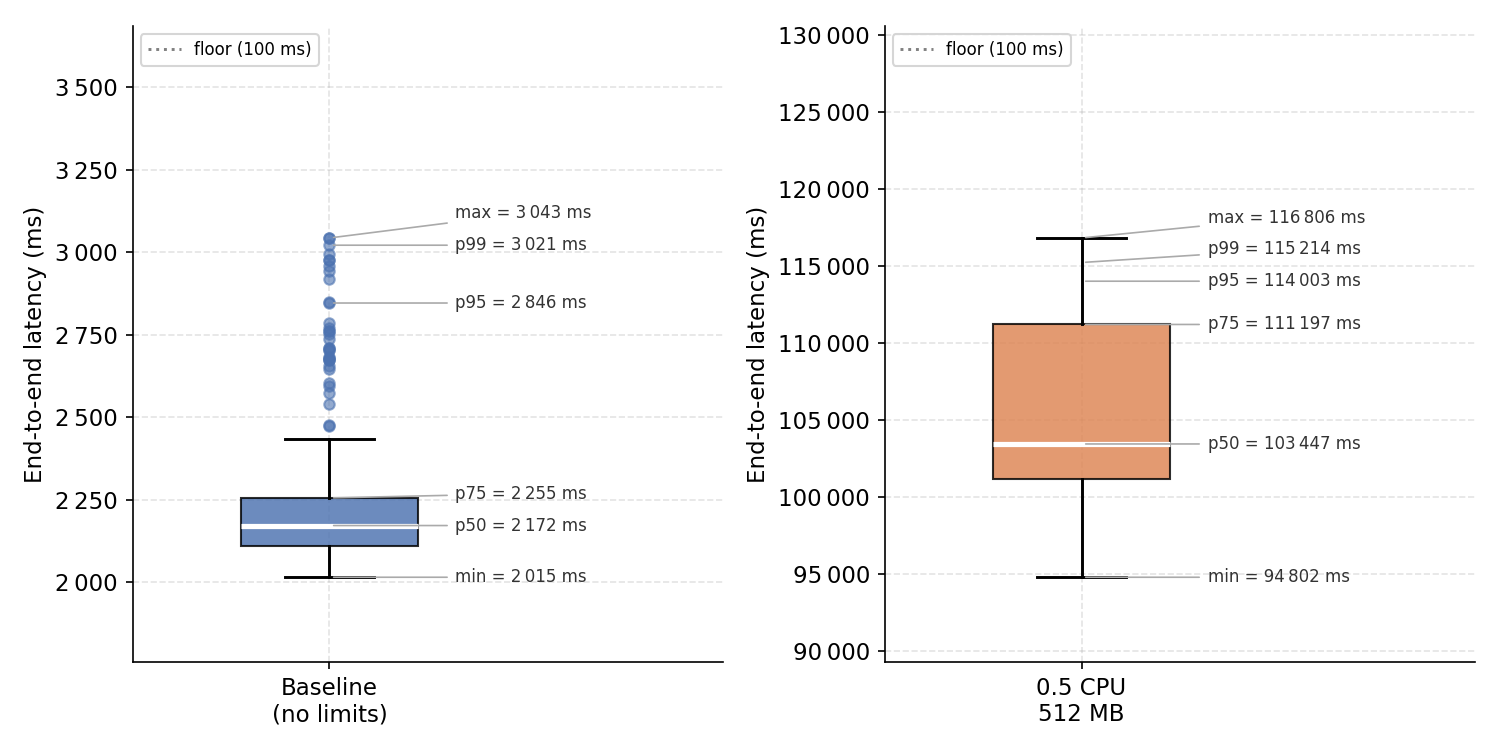}
  \caption{Latency distribution under resource constraints (baseline and 0.5 CPU configurations).}
  \label{fig:rq1_latency}
\end{figure}

Figure~\ref{fig:rq1_latency} presents latency distributions for the two working scenarios. Under baseline conditions, latency ranges from approximately 2,000 ms to just over 3,000 ms. Under the 0.5 CPU constraint, it increases to between 94,000 ms and 116,000 ms, representing a difference of nearly two orders of magnitude.

At baseline, the median (p50) latency was 2,172 ms, with p95 at 2,846 ms and a maximum of 3,043 ms. The minimum observed latency was 2,015 ms, establishing an empirical floor approximately 20$\times$ higher than the 100 ms neuromorphic simulation window. This gap represents the irreducible cost of SNN inference, including Poisson spike encoding, tensor operations in PyTorch, and system-level overhead such as HTTP communication.

When container resources were restricted to 0.5 CPU cores and 512 MB, median latency increased to 103,447 ms, which is a 47.6$\times$ degradation relative to baseline. The p95 reached 114,003 ms and the maximum observed latency was 116,806 ms, corresponding to over 116 seconds per inference request. Unlike the baseline where most requests are close to the median with a small number
higher-latency outliers, the constrained distribution is more dispersed and forms two loose clusters.  This suggests that CPU throttling amplifies input-dependent differences in computational cost, causing certain inputs to consistently exhibit higher latency rather than affecting all requests uniformly.

\begin{figure}[htbp]
  \centering
  \includegraphics[width=0.9\linewidth]{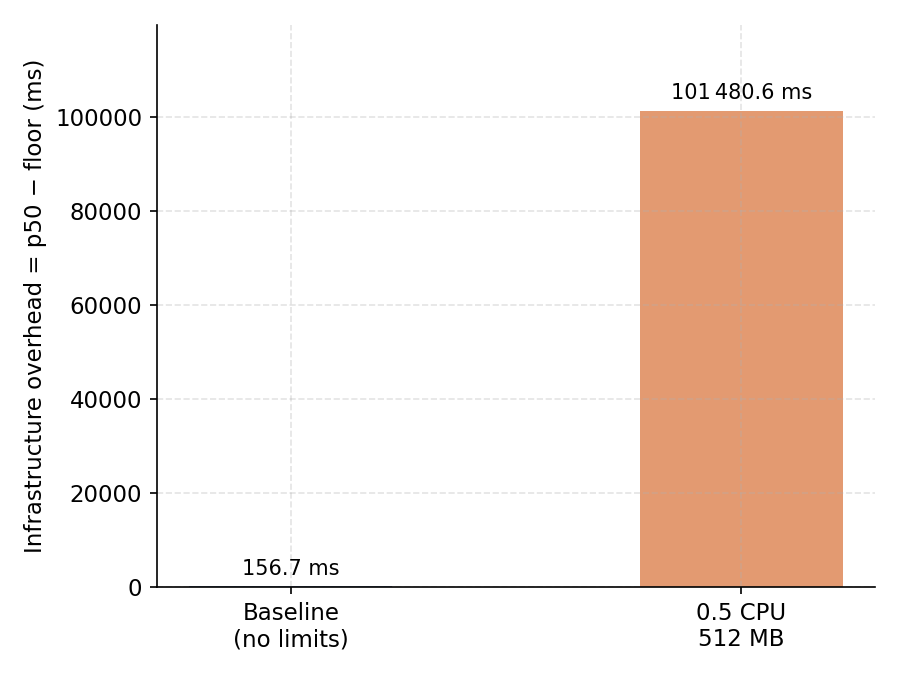}
  \caption{Infrastructure overhead (p50 minus empirical floor) across configurations.}
  \label{fig:rq1_overhead}
\end{figure}

Figure~\ref{fig:rq1_overhead} compares infrastructure overhead across between the scenarios. At baseline, overhead is approximately 157 ms, indicating a small and stable contribution from the container runtime and HTTP stack. Under constrained conditions, overhead increases to 101,481 ms, meaning that approximately 98\% of total latency is attributable to CPU throttling rather than the simulation itself. This is due to the PyTorch-based simulation being computationally intensive, so any restriction in available CPU directly translates into increased execution time. 

\begin{figure}[htbp]
  \centering
  \includegraphics[width=0.9\linewidth]{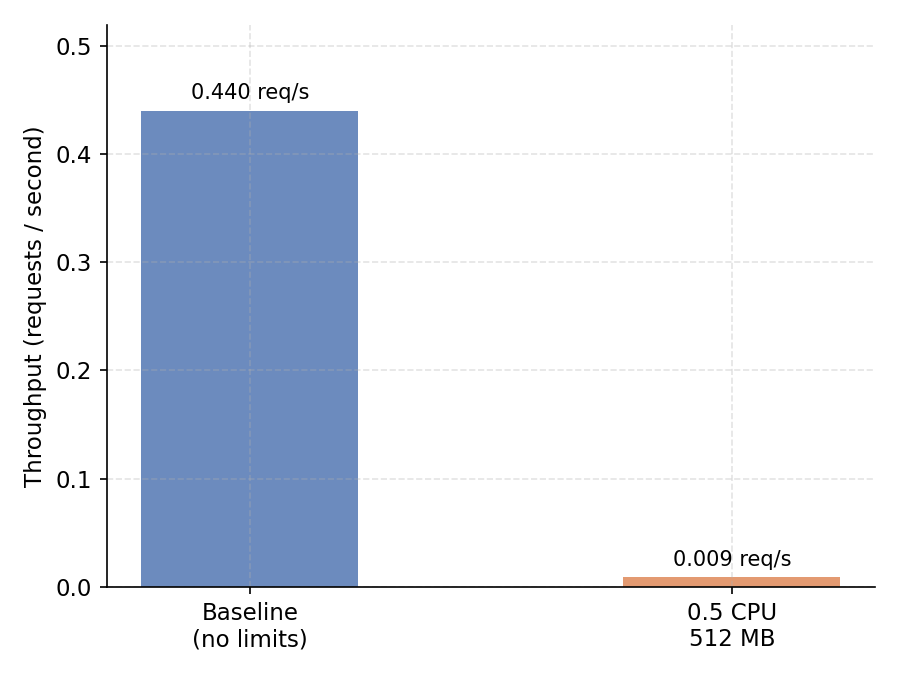}
  \caption{Throughput under resource constraints.}
  \label{fig:rq1_throughput}
\end{figure}

The impact of resource constraints is further reflected in throughput, as shown in Figure~\ref{fig:rq1_throughput}). The baseline sustained 0.44 requests per second, while the constrained configuration achieved only 0.009 requests per second—a 49$\times$ reduction. This highlights how the 0.5 CPU constraint pushes the workload beyond its practical operational limits, especially when no parallelism or batching effects are present. Such a throughput level is impractical for real-time or interactive use cases

\begin{figure}[htbp]
  \centering
  \includegraphics[width=0.9\linewidth]{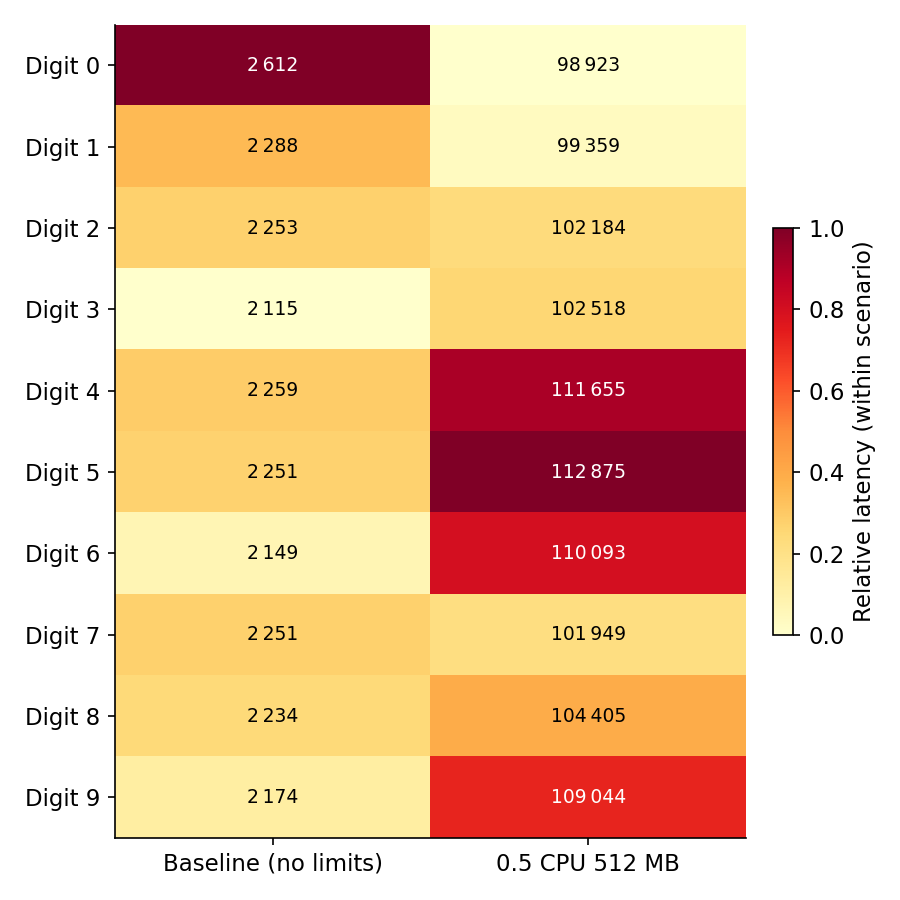}
  \caption{Mean latency per MNIST digit class across configurations.}
  \label{fig:rq1_heatmap}
\end{figure}

Per-digit mean latency is visualized in Figure~\ref{fig:rq1_heatmap}. At baseline, latency is relatively uniform across digit classes. Under constrained conditions, this variation becomes more pronounced and less consistent. Digits 4, 5, and 6 exhibit the highest mean latencies (111,655 to 112,875 ms), while digits 0 and 1 are comparatively lower at approximately 99,000 ms. This suggests that system-level effects, such as scheduling delays and resource contention, have a strong influence on execution time and reduce the predictability of SNN workload behavior

Classification accuracy remained stable across both configurations, with the baseline achieving 76.5\% and 80.5\% under constraint. This confirms that resource
constraints do not degrade the model’s prediction quality, but instead affect only performance. Per-digit accuracy patterns, shown in Table~\ref{tab:accuracy}, are consistent across configurations. Digits 2, 4, 7, and 8 remain more difficult to classify, indicating that these limitations are inherent to the model rather than the deployment environment. 

\renewcommand{\arraystretch}{1.3}
\setlength{\tabcolsep}{8pt}

\begin{table}[htbp]
\centering
\caption{Per-digit classification accuracy across configurations}
\label{tab:accuracy}
\begin{tabular}{|c|c|c|c|}
\hline
\textbf{Digit} & \textbf{Baseline} & \textbf{0.5 CPU} & \textbf{Change} \\
\hline
0 & 95\% & 100\% & +5\% \\\hline
1 & 100\% & 100\% & 0\% \\\hline
2 & 45\% & 55\% & +10\% \\\hline
3 & 90\% & 75\% & $-$15\% \\\hline
4 & 60\% & 60\% & 0\% \\\hline
5 & 90\% & 95\% & +5\% \\\hline
6 & 100\% & 95\% & $-$5\% \\\hline
7 & 65\% & 80\% & +15\% \\\hline
8 & 40\% & 60\% & +20\% \\\hline
9 & 80\% & 85\% & +5\% \\
\hline
\end{tabular}
\end{table}

The most constrained configuration (0.25 CPU / 256 MB) resulted in an OOM failure during container initialization. This means that the PyTorch runtime, BindsNET computation graph, and model weights together require more than 256 MB at startup. This establishes a hard lower bound on the model's resource requirements and is particularly relevant for edge deployment, where many devices provide between 256 MB and 512 MB of RAM.

Table~\ref{tab:rq1_summary} summarizes the key performance metrics across all configurations.

\renewcommand{\arraystretch}{1.3}
\setlength{\tabcolsep}{8pt}

\begin{table}[htbp]
\centering
\caption{RQ1 performance summary}
\label{tab:rq1_summary}
\begin{tabular}{|l|c|c|c|}
\hline
\textbf{Metric} & \textbf{Baseline} & \textbf{0.5 CPU} & \textbf{0.25 CPU} \\
\hline
Throughput (req/s) & 0.44 & 0.009 & OOM \\\hline
Latency p50 (ms) & 2,172 & 103,447 & — \\\hline
Latency p95 (ms) & 2,846 & 114,003 & — \\\hline
Latency max (ms) & 3,043 & 116,806 & — \\\hline
Infra. overhead (ms) & 157 & 101,481 & — \\\hline
Accuracy (proportion) & 76.5\% & 80.5\% & — \\
\hline
\end{tabular}
\end{table}

\subsection{RQ2: K3d Orchestration Under SNN Workloads}

Regarding deployment complexity, K3d provides a straightforward way to implement a single-node virtual edge cluster. A functional K3s cluster can be created with minimal commands, and all components run within Docker containers. However, several practical challenges were encountered. Manual adjustment of the kubeconfig server address was required after cluster creation, and additional steps were needed to connect the development container to the K3d network. Furthermore, the long execution time of SNN inference conflicted with Kubernetes health probes, requiring tuning of probe intervals, timeouts, and failure thresholds in the deployment manifest. Under concurrent load, default Traefik timeout settings resulted in request failures, requiring ingress configuration adjustments to increase proxy timeouts. These issues show that while K3d is operationally simple, its default configuration is designed for fast, stateless services and therefor requires adjustment when supporting longer-running inference workloads such as SNNs.

\begin{figure}[htbp]
  \centering
  \includegraphics[width=\linewidth]{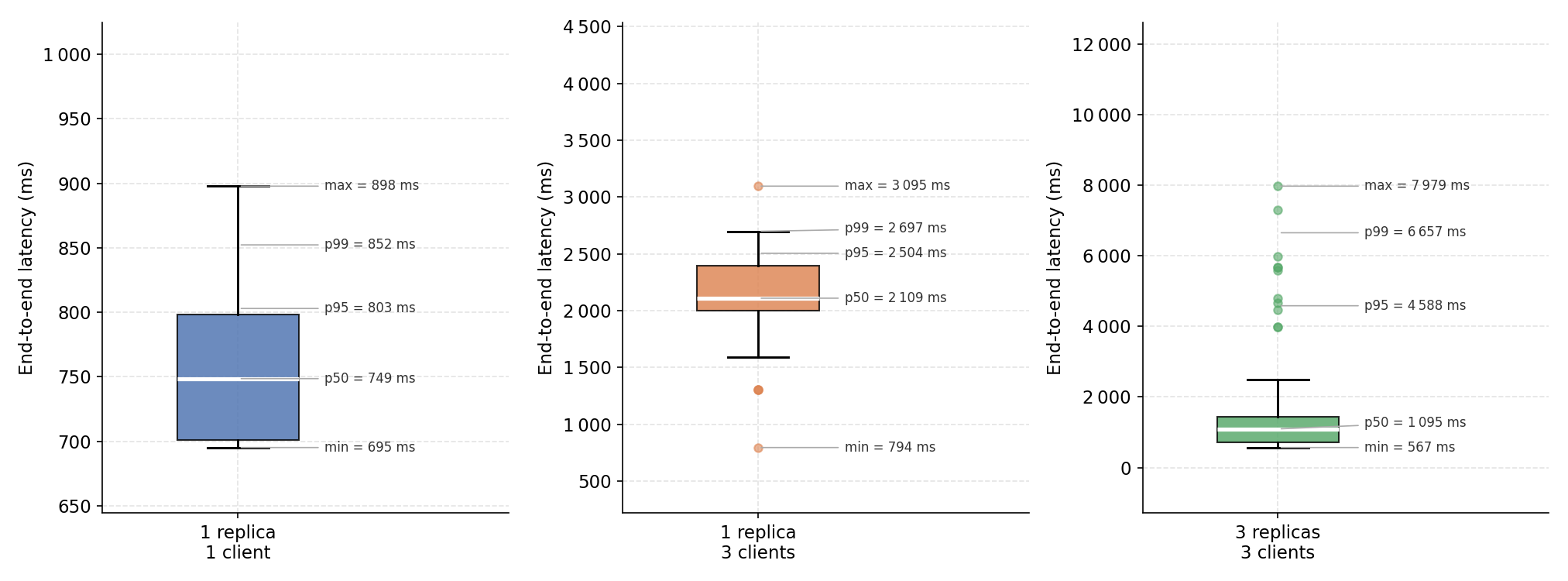}
  \caption{Latency distribution across all three K3d scenarios.}
  \label{fig:rq2_latency}
\end{figure}

Figure~\ref{fig:rq2_latency} presents latency distributions across all three scenarios. Under sequential load (1 replica, 1 client), the system achieved a median latency of 749 ms, a p99 of 852 ms, and a minimum of 695 ms. The distribution is the tightest of all three scenarios,
with a total spread of only 203 ms from minimum to maximum. This suggests that under
no queuing pressure, the performance is highly predictable with minimal orchestration
overhead.

When three clients issue requests concurrently against a single replica, latency increases significantly due to queuing. The median rises to 2,109 ms, a 2.8$\times$ increase, while throughput remains nearly unchanged at approximately 1.37 requests per second. This confirms the system is limited by a single-worker bottleneck, where requests are processed sequentially and additional clients simply queue. The upper tail of the distribution reflects a queue of depth three, where the worst-case request waits approximately two inference periods before its own execution begins.

\begin{figure}[htbp]
  \centering
  \includegraphics[width=\linewidth]{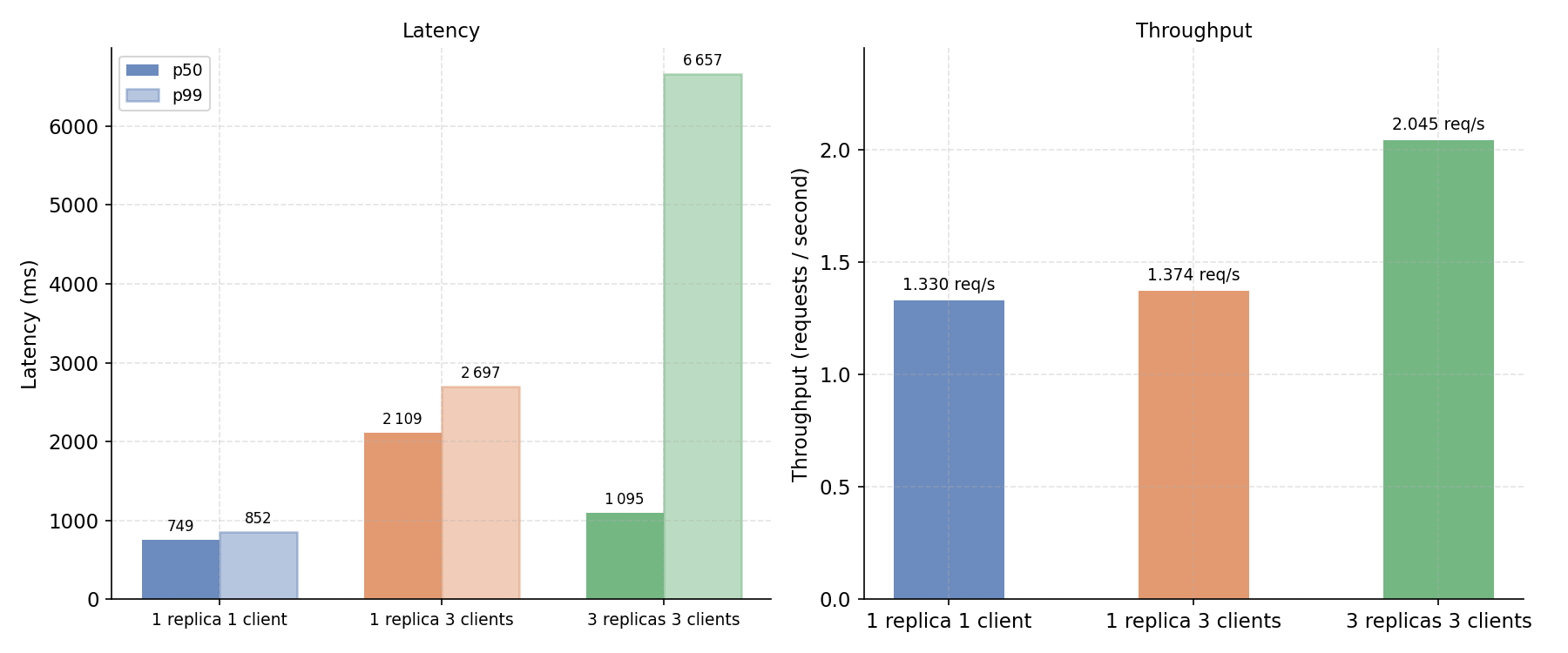}
  \caption{Latency summary and throughput across all three K3d configurations.}
  \label{fig:rq2_scaling}
\end{figure}

Scaling to three replicas while maintaining three concurrent clients reduces median latency to 1,095 ms and increases throughput to 2.05 requests per second. However, this improvement is accompanied by a substantial increase in high-latency requests: p99 rises to 6,657 ms, with maximum latency reaching 7,979 ms (Figure~\ref{fig:rq2_scaling}).

This behavior is caused by the load balancing strategy of the Traefik ingress controller.Traefik employs a round-robin routing policy, which distributes requests evenly without considering the current load on each replica. Because SNN inference takes a substantial amount of time, a pod can still be processing its previous request when the next one arrives via round-robin, causing some replicas to accumulate queues while others remain underutilized.

Throughput scaling efficiency is also below theoretical expectations. With three independent workers each capable of 1.33 requests per second, the theoretical maximum is 3.99 requests per second. The observed result of 2.05 requests per second represents only 51\% of this ideal, meaning the remainder is lost to uneven load distribution.

Table~\ref{tab:rq2_summary} summarizes all three scenarios. The p99/p50 ratio is the most revealing indicator as it shows the growing disparity between typical and worst-case request latency. This ratio grew from 1.1$\times$ under sequential load, to 1.3$\times$ under queuing, and finally to 6.1$\times$ under replica scaling. Even though K3d delivers measurable throughput gains, its default round-robin routing policy produces severe tail latency for long-running inference workloads.

\renewcommand{\arraystretch}{1.3}
\setlength{\tabcolsep}{8pt}

\begin{table}[htbp]
\centering
\caption{RQ2 performance summary across concurrent load scenarios}
\label{tab:rq2_summary}
\begin{tabular}{|l|c|c|c|c|}
\hline
\textbf{Scenario} & \textbf{p50} & \textbf{p99} & \textbf{p99/p50} & \textbf{Req/s} \\
\hline
1 replica / 1 client & 749 & 852 & 1.1$\times$ & 1.33 \\\hline
1 replica / 3 clients & 2,109 & 2,697 & 1.3$\times$ & 1.37 \\\hline
3 replicas / 3 clients & 1,095 & 6,657 & 6.1$\times$ & 2.05 \\
\hline
\end{tabular}
\end{table}

\section{Conclusion}

This paper investigated the feasibility of orchestrating simulated neuromorphic workloads, specifically SNNs, within a virtual edge environment using container-based technologies.

The results demonstrate that SNN inference is highly sensitive to resource availability. Under baseline conditions, the model achieved a median latency of 2,172 ms and a throughput of 0.44 requests per second, reflecting the inherent cost of the simulation. When resources were constrained to 0.5 CPU and 512 MB, latency increased by 47.6$\times$ and throughput decreased by 49$\times$, with the majority of execution time attributable to CPU throttling. The most constrained configuration failed due to insufficient memory, establishing a clear lower bound on resource requirements. Despite these performance changes, classification accuracy remained stable.

From an orchestration perspective, K3d can successfully deploy and scale SNN workloads in a virtual edge setting. Under sequential execution the system exhibited stable and predictable latency. Scaling from one to three replicas improved throughput and reduced median latency, confirming effective parallel execution. However, the default round-robin routing policy introduced load imbalance across replicas, causing some requests to queue while others were processed immediately. This highlights a fundamental mismatch between computation-heavy workloads and the stateless load balancing assumptions of standard container orchestration systems.

Future work can build on these findings in several directions. Exploring alternative load balancing strategies, such as least-connections or occupancy-aware routing, may reduce tail latency for long-running inference workloads. Decoupling inference from the request-response cycle through asynchronous processing could improve responsiveness and resolve conflicts with health checks. Reducing the model size and memory footprint would allow deployment on more constrained edge devices, extending the applicability of SNN workloads. Finally, validation on physical edge hardware is also needed to assess how these results translate to real-world environments with network variability and heterogeneous resources. 

\bibliographystyle{IEEEtran}
\bibliography{references}

\end{document}